\documentclass[preprint,aps,prd,showpacs,superscriptaddress,nofootinbib,tightenlines]{revtex4}

\usepackage{amsfonts}
\usepackage{multirow}
\usepackage{mathrsfs}
\usepackage{graphicx}
\usepackage{amsmath}
\usepackage{amssymb}
\usepackage{bm}
\usepackage{bbm}
\usepackage{color}
\usepackage{slashed}
\usepackage{diagbox}
\usepackage{ulem}
\usepackage{slashbox}
\usepackage{wrapfig}

\def\OMIT#1{}

\newcommand{\nn}{\nonumber}

\newcommand{\beq}{\begin{equation}}
\newcommand{\eeq}{\end{equation}}
\newcommand{\bqa}{\begin{eqnarray}}
\newcommand{\eqa}{\end{eqnarray}}

\begin{document}
\title{\mbox{}\\[11pt]
Mixed electroweak-QCD corrections to $e^+e^-\to \mu^+\mu^- H$ at CEPC
with finite-width effect}


\author{Wen Chen~\footnote{Email: wchen1@ualberta.ca}}
\affiliation{Institute of High Energy Physics, Chinese Academy of Science, Beijing 100049,
China\vspace{0.2cm}}
\affiliation{School of Physics, University of Chinese Academy of
 Sciences, Beijing 100049, China\vspace{0.2cm}}
 \affiliation{Department of Physics, University of Alberta, Canada\vspace{0.2cm}}

\author{Feng Feng~\footnote{Email: F.Feng@outlook.com}}
\affiliation{China University of Mining and Technology, Beijing 100083, China\vspace{0.2cm}}
\affiliation{School of Physics, University of Chinese Academy of
 Sciences, Beijing 100049, China\vspace{0.2cm}}

\author{Yu Jia~\footnote{Email: jiay@ihep.ac.cn}}
\affiliation{Institute of High Energy Physics, Chinese Academy of Science, Beijing 100049,
China\vspace{0.2cm}}
\affiliation{School of Physics, University of Chinese Academy of
 Sciences, Beijing 100049, China\vspace{0.2cm}}

\author{Wen-Long Sang~\footnote{Email: wlsang@ihep.ac.cn}}
 \affiliation{School of Physical Science and Technology, Southwest University, Chongqing 400700, China\vspace{0.2cm}}

\date{\today}

\begin{abstract}
The associated production of Higgs boson with a muon pair, $e^+e^-\to \mu^+\mu^- H$,
is one of the golden channels to pin down the properties of the Higgs boson in the prospective Higgs factories
exemplified by \textsf{CEPC}. The projected accuracy of the corresponding
cross section measurement is about per cent level at CEPC.
In this work, we investigate both ${\mathcal O}(\alpha)$ weak correction
and the ${\mathcal O}(\alpha\alpha_s)$ mixed electroweak-QCD corrections for this channel,
appropriately taking into account the effect of finite $Z^0$ width. The $\mu^+\mu^-$ invariant mass spectrum
is also predicted.
The mixed electroweak-QCD correction turns out to reach 1.5\% of the Born-order result,
and thereby must be included in future confrontation with the data.
We also observe that, after including higher-order corrections,
the simplified prediction for the integrated cross section employing the
narrow-width-approximation may deviate from our full result by a few per cents.
\end{abstract}

\maketitle

\section{Introduction}

Ever since the ground-breaking discovery of the Higgs boson at Large Hadron Collider (\textsf{LHC})
in 2012~\cite{Aad:2012tfa,Chatrchyan:2012xdj}, one of the highest priorities of particle physics
is to nail down the properties of the Higgs boson as precise as possible.
Unlike the hadron colliders, which suffer from severe contamination due to the copious background events,
the electron-positron colliders provide an ideal platform to precisely measure various Higgs couplings~\cite{Accomando:1997wt}.
In recent years, three next-generation $e^+e^-$ colliders have been proposed for dedicated study of Higgs boson:
International Linear Collider (ILC)~\cite{Baer:2013cma},
Future Circular Collider (FCC-ee)~\cite{Gomez-Ceballos:2013zzn},
and Circular Electron-Positron Collider (CEPC)~\cite{CEPC-SPPCStudyGroup:2015csa},
all of which plan to operate at the center-of-mass energy around $240\sim 250$ GeV.

There emerge several Higgs production mechanisms at $e^+e^-$ colliders: Higgsstrahlung, $WW$ fusion and $ZZ$ fusion, {\it etc.}.
Around $\sqrt{s}\approx 240$ GeV, which is the projected energy range of \textsf{CEPC}, the Higgs production is dominated by the
Higgsstrahlung channel $e^+e^-\to Z H$, Higgs production associated with a $Z^0$ boson.
It is anticipated that, with the aid of very high luminosity and the recoil mass technique, \textsf{CEPC} can measure the Higgs production cross section with an exquisite sub-per-cent accuracy.
Needless to say, it is indispensable for theoretical predictions for the Higgsstrahlung channel to be
commensurate with the projected experimental precision.

The leading order (LO) prediction for $e^+e^-\to ZH$ was first considered in 70s~\cite{Ellis:1975ap,Lee:1977eg,Lee:1977yc,Ioffe:1976sd}.
In the early 90s, the next-to-leading order (NLO) electroweak correction for this process has also been addressed by three groups independently~\cite{Fleischer:1982af,Kniehl:1991hk,Denner:1992bc}, which turns out to be significant.
Very recently, the mixed electroweak-QCD next-to-next-to-leading (NNLO) corrections were also be independently calculated
by two groups \cite{Gong:2016jys,Sun:2016bel}. The ${\cal O}(\alpha\alpha_s)$ correction may reach 1\% of the LO prediction,
thereby must be included when confronting the future measurement. Recently, the ISR effect of this process has also been carefully analyzed~\cite{Mo:2015mza}.

From the experimental angle, it is the decay products of the $Z^0$ boson, rather than the $Z^0$ itself
that are tagged by detectors in the Higgsstrahlung channel, since the $Z^0$ is an unstable particle.
Therefore, in order to get closer contact with experiment, it is advantageous to make precise
predictions directly for the process $e^+e^-\to (Z^{*}\to) f\bar{f}+H$,
where $f$ represents leptons or quarks. Among a flurry of Higgs production channels
associated with various $Z$ decay products, the $e^+e^-\to\mu^+\mu^-H$ process occupies a unique place for probing Higgs properties,
because it is a very clean channel and possesses large cross section.
The production cross section for this individual channel can be measured with 0.9\% precision at \textsf{CEPC}~\cite{CEPC-SPPCStudyGroup:2015csa,Chen:2016zpw}.
Combining several other channels, \textsf{CEPC} is anticipated to measure the Higgs production rate
with the accuracy of $0.51$\%.

The LO contribution to the $e^+e^-\to f\bar{f}+H$ process was first considered in 70s~\cite{Jones:1979bq}.
The initial-state-radiation (ISR) correction to these types of processes was addressed in 80s~\cite{Berends:1984dw}.
There exist a flurry of higher-order studies for the process $e^+e^-\to\nu\bar{\nu}H$,
where both Higgsstrahlung and $WW$ fusion mechanisms contribute~\cite{Altarelli:1987ue,
Boos:1993uf,Patrignani:2016xqp,Belanger:2002me,Denner:2003yg,Denner:2003iy,Belanger:2003sd,Denner:2004jy}.
To our knowledge, there appears no dedicated work to investigate the NLO weak correction to $e^+e^-\to\mu^+\mu^-H$.
Nevertheless, the NLO weak correction to a similar process $e^+e^-\to e^+e^-H$ were calculated by the
\textsf{GRACE} group more than a decade ago~\cite{Belanger:2002ik,Boudjema:2004eb}.
One can extract the corresponding NLO weak correction to $e^+e^-\to\mu^+\mu^-H$
by singling out a subset of diagrams in \cite{Belanger:2002ik,Boudjema:2004eb}.

The purpose of this work is to conduct a systematic investigation on the higher-order
radiative corrections to the process
$e^+e^-\to\mu^+\mu^-H$, to match the projected experimental precision at \textsf{CEPC}.
We first compute the NLO weak correction to $e^+e^-\to\mu^+\mu^-H$, then proceed to
include the ${\mathcal O}(\alpha\alpha_s)$ mixed electroweak-QCD correction.
Besides the integrated cross section, we also study the impact of radiative corrections to
various kinematic distributions such as the $\mu^+\mu^-$ invariant mass distribution.
For this purpose, the finite $Z^0$ width effect must be consistently taken into account.
It is also instructive to examine how our results deviate from those obtained by
invoking the narrow width approximation (NWA).

The rest of the paper is structured as follows.
In section~\ref{LO:NWA}, adopting the Breit-Wigner ansatz for the resonant $Z^0$ propagator,
we recapitulate the LO prediction for $e^+e^-\to\mu^+\mu^-H$ and also show the corresponding NWA result.
In section~\ref{Treatment:Z:width}, we specify our strategy of
implementing the finite $Z^0$-width effect in higher-order calculation.
In section~\ref{Sec:NLO:Weak:Correction}, we present the calculation for the NLO weak correction to this channel.
In section~\ref{Sec:NNLO:mixed:weak:QCD:Correction}, we describe the calculation for the mixed electroweak-QCD corrections.
In section~\ref{Numerics}, we present the numerical results and phenomenological analysis.
Finally we summarize in section~\ref{Conclusion}.

\section{Leading order results and narrow width approximation}
\label{LO:NWA}

 \begin{figure}[htbp]
 	\centering
 	\includegraphics[width=0.4\textwidth]{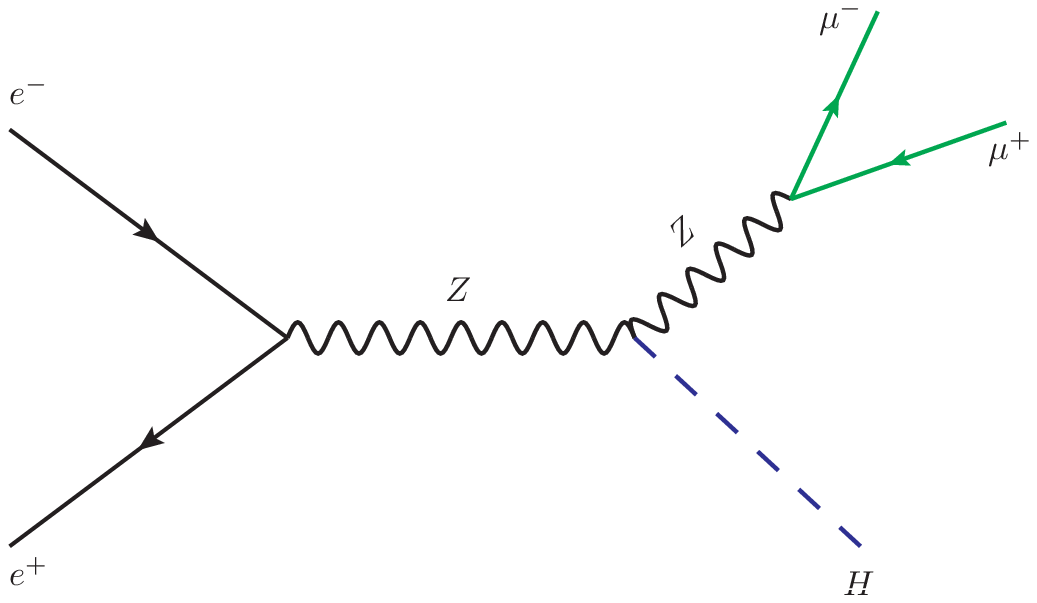}
 	\caption{LO diagram for $e^+e^-\to \mu^+\mu^-H$.
 \label{Fig:LO:diagrams}}
 \end{figure}

We are considering the process
\beq
e^+(k_1)+e^-(k_2)\to \mu^+(p_1)+\mu^-(p_2)+H(p_H),
\eeq
where the momenta of the incoming and outgoing particles are specified in the parentheses.
For future usage, we define $s\equiv(k_1+ k_2)^2$,
and $s_{12}\equiv (p_1+p_2)^2$. For convenience, we also define the invariant
mass of the muon pair by $M_{\mu\mu}\equiv \sqrt{s_{12}}$, which lies in the range $0\le M_{\mu\mu} \le \sqrt{s}-M_H$.

At Higgs factory, lepton masses can be safely neglected owing to their exceedingly small Yukawa couplings.
Consequently at the lowest order, there is only a single $s$-channel diagram as depicted in Fig.~\ref{Fig:LO:diagrams}.
The LO amplitude reads
\bqa
\label{LO:amplitude}
\widetilde{\mathcal M}_0  &=&  -{e^3M_Z \over s_W c_W} \bar{v}(k_1)\Gamma^{\mu}_Z  u(k_2) {g_{\mu\nu}\over  (s-M_Z^2) ( s_{12}-M_Z^2)}
 \bar{u}(p_1)\Gamma_{Z}^\nu v(p_2),
\eqa
where $c_W\equiv \cos \theta_W$, $s_W\equiv \sin \theta_W$, with $\theta_W$ the Weinberg angle, $M_Z$
represents the mass of the $Z^0$ boson.
$\Gamma^{\mu}_V= g_V^+\gamma^{\mu}\frac{1+\gamma^5}{2}+ g_V^-\gamma^{\mu}\frac{1-\gamma^5}{2}$
is the coupling of the gauge boson and the charged lepton.
Specifically speaking, $g_Z^+=\frac{s_W}{c_W}$, $g_Z^-=\frac{s_W}{c_W}-\frac{1}{2s_Wc_W}$.
The chirality structure of the neutral current demands that, $e^+$ and $e^-$ (also $\mu^+$ and $\mu^-$)
must carry opposite helicity in order to render a non-vanishing amplitude.

As can be readily seen from Fig.~\ref{Fig:LO:diagrams}, it is possible for the
$\mu^+\mu^-$ pair to be resonantly produced from the on-shell $Z^0$ boson,
consequently the amplitude in \eqref{LO:amplitude} blows up at $s_{12}=M_Z^2$,
which reflects that fixed-order calculation breaks down near the $Z$ pole.
To tame the singularity in the limit $s_{12} \to M_Z^2$,
it is customary to replace the second $Z$ boson propagator in (\ref{LO:amplitude}) with the
Breit-Wigner form, which amounts to include the Dyson summation for the $Z$ boson self-energy diagrams.
Retaining finite $Z$ width would effectively cutoff the IR singularity.
One may define a new amplitude:
\beq
\mathcal{M}_0= \mathcal{F} \widetilde{\mathcal M}_0, \qquad\qquad
\mathcal{F} = \frac{s_{12}-M_Z^2}{s_{12}-M_Z^2+iM_Z\Gamma_Z},
\label{rescaling:formula}
\eeq
where $\cal F$ is a rescaling factor, and $\Gamma_Z$ signifies the width of the $Z^0$ boson.

The LO cross section is then given by
\beq
\sigma_0 = {1\over 2s} \int\!\!d\Pi_3\,
{1\over 4}\sum_{\rm Pol}\left|{\mathcal M}_0 \right|^2,
\label{Cross:Section:Definition}
\eeq
where the three-body phase space in the center-of-mass (CM) frame
can be conveniently parameterized as
\bqa
 \int\!\! d \Pi_3 &=& \int \!\! {d^3 p_1 \over (2\pi)^3 2 p_1^0}{d^3 p_2 \over (2\pi)^3 2 p_2^0} {d^3 p_H \over (2\pi)^3 2 p_H^0}
 (2\pi)^4\delta^{(4)}(k_1+k_2-p_1-p_2-p_H)
\nn\\
&=& {1\over (2\pi)^4} {1\over 16 \sqrt{s}} \int \!\! \frac{d s_{12}}{\sqrt{s_{12}}} \, d\Omega_1^*
d \cos \theta_H \, |{\bf p}_1^*| \, |{\bf p}_H|,
\label{three:body:phase:space}
\eqa
where $(|{\bf p}_1^*|, \Omega_1^*)$ signifies the 3-momentum of the $\mu^-$ in the rest frame of
the dimuon system, $|{\bf p}_H|$, $\theta_H$ represent the magnitude of the momentum and the polar angle
of the Higgs boson in the laboratory frame, respectively.
Upon neglecting masses of the electron and muon, one obtains $|{\bf p}_1^*|=M_{\mu\mu}/2$,
and $|{\bf p}_H|= {1\over 2\sqrt{s}}\lambda^{1/2}(s, s_{12},M_H^2)$,
where $\lambda(a,b,c)\equiv a^2+b^2+c^2-2 a b-2 a c -2 b c$ is the K\"{a}ll\'{e}n function.
In deriving (\ref{three:body:phase:space}), we have utilized the axial symmetry to
eliminate the trivial dependence on the azimuthal angle of the outgoing Higgs boson.

Squaring (\ref{LO:amplitude}), summing over $\mu^+\mu^-$ helicities,
and averaging upon the $e^+ e^-$ polarizations,
one observes that the squared amplitude bears a factorized structure,
thanks to the simple $s$-channel topology.
Substituting it into (\ref{Cross:Section:Definition}), integrating over the solid angle $\Omega_1^*$,
one then arrives at the following double differential cross section:
\begin{align}
& {d^2 \sigma_0 \over d s_{12} d\cos \theta_H}= {\alpha^3 \left(g^{+2}_Z + g^{-2}_Z \right)^2 \over 24 c_W^2s_W^2}
 {|{\bf p}_H| M_Z^2 \over  \sqrt{s} \left(s-M_Z^2\right)^2}
 {s_{12}\over (s_{12}-M_Z^2)^2+M_Z^2 \Gamma_Z^2}\left(2 + \sin^2\theta_H {{\bf p}_H^2\over s_{12}}\right),
\label{double:diff:X:Section:Born:Order}
\end{align}
with $\alpha \equiv {e^2\over 4\pi}$ the electromagnetic fine structure constant.

Integrating (\ref{double:diff:X:Section:Born:Order}) over the polar angle,
one obtains the Born-order spectrum of the invariant mass of $\mu^+\mu^-$:
\bqa
&& {d \sigma_0 \over d M_{\mu\mu} }=
{\alpha^3 \left(g^{+2}_Z + g^{-2}_Z \right)^2 \over 9 c_W^2s_W^2}
 {|{\bf p}_H| M_Z^2 \over  \sqrt{s} \left(s-m_Z^2\right)^2}
 { s_{12}^{3/2} \over (s_{12}-M_Z^2)^2+M_Z^2 \Gamma_Z^2} \left(3 +  {{\bf p}_H^2\over s_{12}}\right).
\label{inv:mass:distr:Born:Order}
\eqa

Since $\Gamma_Z\ll M_Z$, one naturally expects that the NWA
should be fairly reliable for the process under consideration.
Inserting the limiting formula
\beq
\lim_{\Gamma_Z\to 0} {1\over (s_{12}-M_Z^2)^2+M_Z^2 \Gamma_Z^2} = {\pi\over M_Z \Gamma_Z} \delta(s_{12}-M_Z^2)
\label{NWA:formula}
\eeq
into (\ref{double:diff:X:Section:Born:Order}), and integrating over $s_{12}$,
we obtain the angular distribution:
\bqa
&& {d \sigma_0 \over d\cos \theta_H}\Big |_{\rm NWA}= {d \sigma_0(ZH) \over d \cos\theta} {\rm Br}_0
(Z\to \mu^+\mu^-),
\label{angle:distr:Born:Order:NWA}
\eqa
where
\beq
\label{LO:ZH:Differential:X:section}
{d \sigma_0(ZH) \over d \cos\theta} = {\pi \alpha^2 \left(g^{+2}_Z + g^{-2}_Z \right) \over 4 c_W^2 s_W^2}
{|{\bf p}_H| M_Z^2 \over \sqrt{s} (s-M_Z^2)^2} { \over }  \left( 2+ \sin^2\theta {{\bf p}_Z^2 \over M_Z^2} \right),
\eeq
is the angular distribution of the $Z$($H$) in the process $e^+e^-\to ZH$ at Born order,
with $|{\bf p}_H| \equiv {1\over 2\sqrt{s}}\lambda^{1/2}(s, M_Z^2, M_H^2)$.
In (\ref{angle:distr:Born:Order:NWA}), the Born-order partial width and branching fraction
of $Z\to \mu^+\mu^-$ are given by
\begin{subequations}
\bqa
&& \Gamma_0(Z\to \mu^+\mu^-)={\alpha\over 6}\left(g^{+2}_Z + g^{-2}_Z \right) M_Z,
\\
&& {\rm Br}_0(Z\to \mu^+\mu^-) \equiv {\Gamma_0(Z\to \mu^+\mu^-)\over \Gamma_Z}.
\eqa
\label{Born:Order:Z:mumu:partial:width:Bran:Frac}
\end{subequations}

From (\ref{angle:distr:Born:Order:NWA}), one readily obtains the
LO integrated cross section in the NWA ansatz:
\bqa
&& \sigma_0 (\mu^+\mu^- H)\Big |_{\rm NWA}= \sigma_0(ZH) {\rm Br}_0 (Z\to \mu^+\mu^-),
\label{Integrated:X:Section:Born:Order:NWA}
\eqa
where
\bqa
\label{Unpol:LO:ZH:X:section}
&& \sigma_0(ZH)=\frac{\pi\alpha^2\left(g^{+2}_Z + g^{-2}_Z \right)}{3c_W^2 s_W^2}{|{\bf p}_H|M_Z^2\over\sqrt{s}
(s-M_Z^2)^2}\left(3+ {{\bf p}_Z^2 \over M_Z^2}\right).
\eqa
Note that the unpolarized LO cross section $\sigma_0(ZH)$ in (\ref{Unpol:LO:ZH:X:section})
decreases rather mildly ($\propto 1/s$) in the high energy limit, reflecting
the dominance of producing the longitudinally polarized $Z$ in large $\sqrt{s}$.
However, at moderate energy such as $\sqrt{s}= 250$ GeV at \textsf{CEPC},
the longitudinally-polarized cross section
only comprises of 42\% of the total unpolarized cross section.

\section{The treatment of finite $Z^0$ width in higher-order corrections}
\label{Treatment:Z:width}

As mentioned before, in this work we are interested in addressing the NLO weak and mixed electroweak-QCD
corrections for $e^+e^-\to\mu^+\mu^-H$:
\beq
\label{LOR:eq1}
\mathcal{M}=\mathcal{M}_{0}+\mathcal{M}^{(\alpha)}+\mathcal{M}^{(\alpha\alpha_s)}+\cdots.
\eeq
For simplicity, in this work we have neglected the pure QED corrections (such as ISR and FSR effect),
which can instead be simulated by the package \textsf{Whizard}~\cite{Kilian:2007gr}.
As a consequence, a simplifying feature arises that the dominant higher-order diagrams resemble
the $s$-channel topology as depicted in Fig.~\ref{Fig:LO:diagrams},
which contains only one resonant $Z$ propagator.

Once going beyond LO, it becomes a quite delicate issue to incorporate the finite $Z$ width effect
yet without spoiling gauge invariance and bringing double counting.
Over the past decades, numerous practical schemes have been proposed to tackle the unstable particle, such as
the pole scheme~\cite{Veltman:1963th,Stuart:1991xk,Sirlin:1991fd},
factorization scheme~\cite{Baur:1991pp,Kurihara:1994fz},
fermion-loop scheme~\cite{Beenakker:1996kt,Beenakker:1996kn}, boson-loop scheme~\cite{Beuthe:1996fe},
complex mass scheme~\cite{Denner:1999gp,Denner:2006ic}, {\it etc.}.
It is worth mentioning that a systematic and model-independent approach, the
unstable particle effective theory, has also emerged finally~\cite{Beneke:2003xh,Beneke:2015vfa}.
However, this approach is valid only near the resonance peak, and cannot be applied
in the entire kinematic range.

Owing to the particularly simple $s$-channel topology of our process, it is most convenient to employ the factorization scheme~\cite{Baur:1991pp,Kurihara:1994fz}, which is particularly suitable for such resonance-dominated process.
In this scheme, one rescales a gauge-invariant higher-order amplitude by a Breit-Wigner factor ${\cal F}$,
and subtracting the $iM_Z\Gamma_Z$ terms which potentially generates double counting.
The merit of this scheme is that gauge invariance is preserved, and can be readily
implemented in automated calculation.
Recently this scheme has also been used by Denner {\it et al.} to
analyze the NLO electroweak correction to $e^+e^-\to \nu\bar{\nu}H$~\cite{Denner:2003iy}.

For our purpose, we specify the recipe of the factorization scheme
closely following~\cite{Denner:2003iy}:
\beq
\label{Factorization:scheme}
{\mathcal M}^{(\alpha\alpha_s^n)} = {\mathcal F} \widetilde {\mathcal M}^{(\alpha\alpha_s^n)} +
i\frac{\text{Im}\left\{\hat{\Sigma}_{T}^{ZZ\;(\alpha\alpha_s^n)}(M_Z^2)\right\}}
{s_{12}-M_Z^2}\mathcal{M}_0,
\eeq
where $n=0,1$, $\widetilde{M}$ represents the fixed-order amplitude where the $Z^0$ is treated as a rigorously
stable particle, $\cal F$ and ${\cal M}_0$ have been defined in \eqref{rescaling:formula},
$\hat{\Sigma}^T_{ZZ}(s)$ represents the transverse part of the renormalized one-particle irreducible self-energy diagrams for $Z$ boson.
$M_Z$ is the pole mass of the $Z^0$ boson,
and throughout the work we take $\Gamma_Z$ as the experimentally determined $Z^0$ boson width~\footnote{By default, the pole mass of the $Z^0$
is determined by the condition $\text{Re}\left\{\hat{\Sigma}_{ZZ}(M_Z^2)\right\}\equiv 0$, whereas
its width is inferred from the optical theorem, $M_Z\Gamma_Z={\rm Im}\left\{\hat{\Sigma}_{ZZ}(M_Z^2)\right\}$.
It was argued~\cite{Willenbrock:1991hu,Stuart:1991xk,Sirlin:1991fd} that the pole mass and the corresponding width
defined this way are gauge dependent. Nevertheless, the gauge-dependent terms arise at order-$\alpha^3$
in the 't Hooft-Feynman gauge, which is beyond the accuracy targeted in this work.
Thus we will pretend $M_Z$ and $\Gamma_Z$ to be gauge-invariant quantities.}.

The second term in the right-hand side of \eqref{Factorization:scheme} is included to
subtract the double-counting term. Fortunately, due to its orthogonal phase, the interference of this term with
$\widetilde{\mathcal M}$ generates a purely imaginary contribution to the cross section,
thus can be safely neglected.

Once the rescaled ${\cal O}(\alpha)$ and ${\cal O}(\alpha\alpha_s)$ amplitudes are obtained,
we then deduce the corresponding higher-order corrections to the differential cross section through
\beq
\sigma^{(\alpha\alpha_s^n)} = {1\over 2s} \int\!\!d\Pi_3\,
{1\over 4}\sum_{\rm Pol} 2{\rm Re}\left[{\mathcal M}_0^* {\mathcal M}^{(\alpha\alpha_s^n)}\right],
\label{Higher:order:X:Section:interference}
\eeq
with $n=0,1$. Note even for the mixed electroweak-QCD correction,
we only need consider its interference with the Born-order amplitude.

We conclude this section by stressing that, since the non-resonant diagrams are regular at $s_{12}=M_Z^2$,
the rescaling procedure in \eqref{Factorization:scheme} enforces their contributions to the
amplitude to vanish on the $Z^0$ pole.
In the vicinity of the resonance, it is intuitively appealing that
the non-resonant diagrams are much more suppressed relative to the resonant diagrams.
As will be seen in section~\ref{Numerics}, our numerical predictions indeed confirm this anticipation.

\section{Calculation of the NLO weak correction}
\label{Sec:NLO:Weak:Correction}

\begin{figure}[htbp]
 	\centering
 	\includegraphics[width=1.0\textwidth]{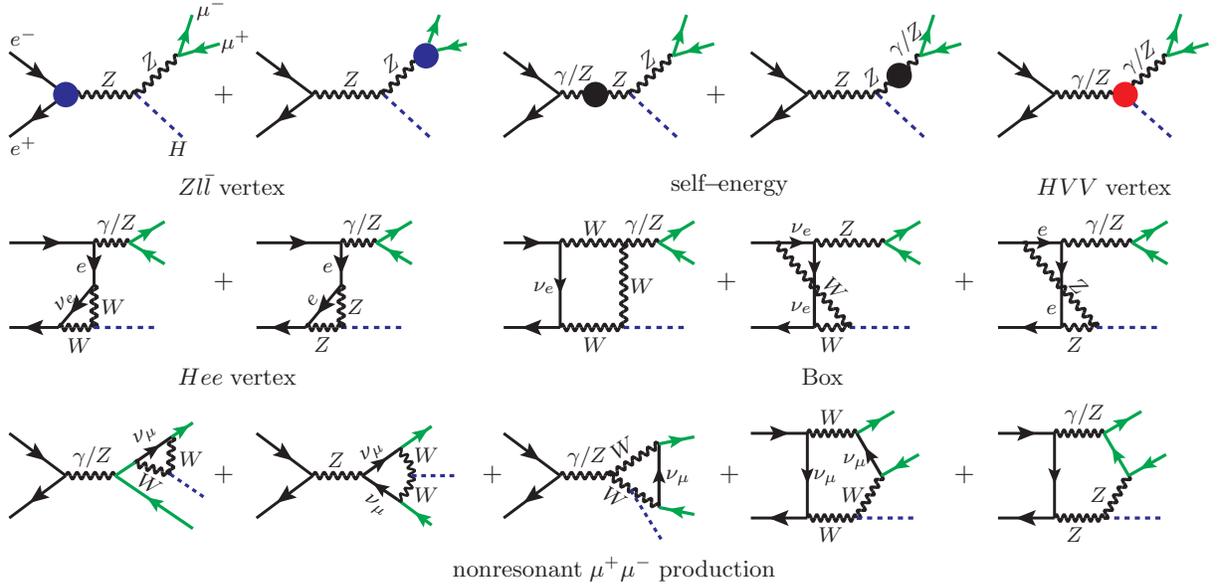}
 	\caption{Some representative higher-order diagrams for $e^+e^-\to \mu^+\mu^-H$,
 	through the order-$\alpha\alpha_s$. The three solid heavy dots are explained in Fig.~\ref{Fig:SE:Vertex:diagrams}.
 Diagrams in the first two rows correspond to the ``resonant'' channel $e^+e^-\to (Z^*/\gamma^*\to) \mu^+\mu^-+H$,
 while those in the last row exhibit a completely different ``non-resonant" topology.
 \label{Fig:eemumuH:higher:order}}
 \end{figure}

\begin{figure}[htbp]
\centering
\includegraphics[width=1.0\textwidth]{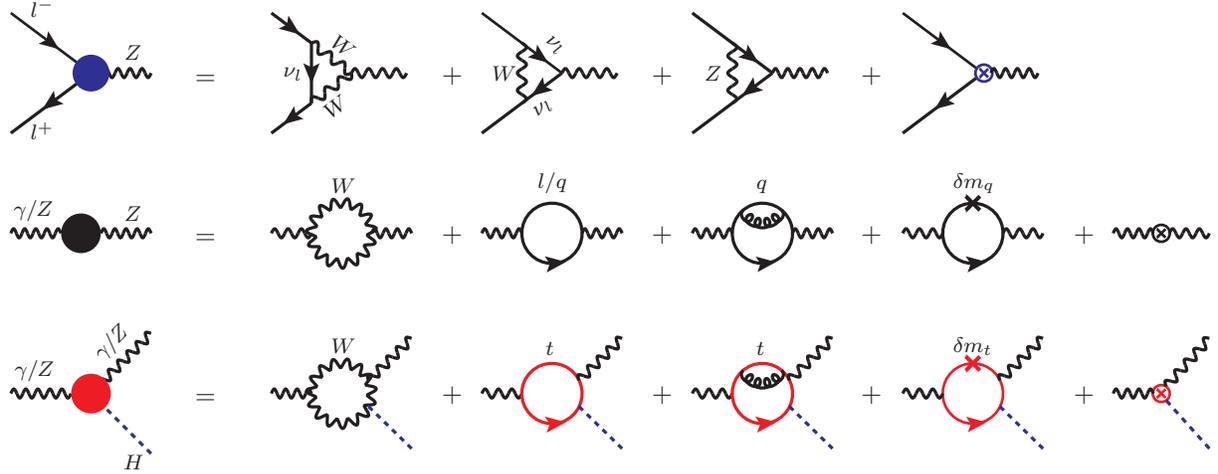}
\caption{Representative diagrams for the radiative corrections to the
renormalized $Zee$ vertex, $\gamma/Z$ self-energy, and $HVV$ vertex, through order-$\alpha\alpha_s$.
The cross represents the
quark mass counterterm in QCD, cap denotes the electroweak counterterm in on-shell scheme.
\label{Fig:SE:Vertex:diagrams}}
\end{figure}

We now outline the calculation of the NLO weak correction to $e^+e^-\to \mu^+\mu^-H$, with some
representative diagrams depicted in Fig.~\ref{Fig:eemumuH:higher:order} and \ref{Fig:SE:Vertex:diagrams}.
As stressed before, we will not consider the ISR and FSR types of diagrams.
It is obvious that the NLO diagrams can be separated into two gauge-invariant subgroups,
with either ``resonant'' or ``non-resonant'' structures.
For the former subset, the diagrams are very similar to
those encountered in the previous NLO weak correction for $e^+e^-\to ZH$, so are the corresponding calculations;
for the latter, there emerges no singularity as $s_{12}\to M_Z^2$,
so there is no need to include width effect for any particle routing around the loop.

The NLO amplitude is computed in Feynman gauge.
Masses of all light fermions are neglected except the top quark.
Dimensional regularization (DR) is employed to regularize UV divergence.
The Feynman diagrams and the corresponding amplitude are generated by the package \textsf{FeynArts}~\cite{Hahn:2000kx}.
Tensor contraction and Dirac/color matrices trace are conducted by using \textsf{FeynCalc} and \textsf{FeynCalcFormLink}~\cite{Mertig:1990an,Feng:2012tk,Shtabovenko:2016sxi}.
Tensor integrals are further reduced to the Passarino-Veltman scalar functions,
which are numerically evaluated by \textsf{Collier}~\cite{Denner:2016kdg}
and \textsf{LoopTools}~\cite{Hahn:1998yk}.

We also choose to use the standard on-shell renormalization scheme to sweep UV divergences,
where various electroweak counterterms are tabulated in \cite{Denner:1991kt}.
Depending on the specific recipe for the charge renormalization constant $Z_e$,
there are three popular sub-schemes of the on-shell renormalization:
$\alpha(0)$, $\alpha(M_Z)$ and $G_\mu$ schemes~\cite{Denner:1992bc}. In the first scheme,
the fine structure constant $\alpha$ is assuming its Thomson-limit value,
whereas $\alpha(0)$ is replaced with
\begin{subequations}
\bqa
\label{alphaMZ:Gmu:scheme}
\alpha(M_Z) &=& \frac{\alpha(0)}{1-\Delta\alpha(M_Z)},
\\
\alpha_{G_\mu} &=& \frac{\sqrt{2}}{\pi}G_{\mu} M_W^2 s_W^2,
\eqa
\end{subequations}
in the $\alpha(M_Z)$ and $G_\mu$ schemes, respectively.
Differing from the $\alpha(0)$ scheme,
these two schemes effectively resum either some universal large
logarithms from the light fermion loop or
some $m_t^2$-enhanced terms from the top quark loop.

Once the $\widetilde{\cal M}^{(\alpha)}$ is rendered finite after the renormalization procedure,
we then employ \eqref{Factorization:scheme} to obtain the rescaled amplitude ${\mathcal M}^{(\alpha)}$,
which encapsulates the finite $Z$-width effect.
It is then straightforward to utilize \eqref{Higher:order:X:Section:interference}
to infer the NLO weak correction to the differential cross section.

\section{Calculation of Mixed Electroweak-QCD Corrections}
\label{Sec:NNLO:mixed:weak:QCD:Correction}

Finally we turn to the ${\cal O}(\alpha\alpha_s)$ mixed electroweak-QCD correction
to $e^+e^-\to \mu^+\mu^-H$. Since it is the quarks instead of leptons that can experience
the strong color force, we only need retain those diagrams involving quark loop.
Moreover, since the top quark couples the Higgs boson with the strongest strength, for simplicity we
have neglected the masses of all lighter quarks,
so we only retain those two-loop diagrams where only the top quark loop dressed by gluon.
Some typical two-loop diagrams are shown in Fig.~\ref{Fig:eemumuH:higher:order} and \ref{Fig:SE:Vertex:diagrams},
bearing only the $s$-channel ``resonant'' structure.
As indicated in Fig.~\ref{Fig:SE:Vertex:diagrams}, at this order, QCD renormalization is
realized by merely inserting the one-loop top quark mass counterterm, $\delta m_t$,
into the internal top-quark propagator, as well as into the $Ht\bar{t}$ vertex~\cite{Sun:2016bel}.
The calculation very much resembles our preceding work on ${\cal O}(\alpha\alpha_s)$ correction to
$e^+e^-\to ZH$~\cite{Sun:2016bel}, and we referred the interested readers
to that paper for more details.

For the actual two-loop computation, we utilize the packages \textsf{Apart}~\cite{Feng:2012iq}
and \textsf{FIRE}~\cite{Smirnov:2014hma} to perform partial fraction
and integration-by-parts (IBP) reduction. We then combine \textsf{FIESTA}~\cite{Smirnov:2015mct}/\textsf{CubPack}~\cite{CubPack} to perform sector decomposition and subsequent numerical integrations for master integrals with quadruple precision.

Besides the finite renormalization of $Zee$ vertex~\cite{Sun:2016bel},
the ${\cal O}(\alpha\alpha_s)$ amplitude can be expressed in terms of the
Born-order amplitude supplemented with an effective $HVV$ vertex:
\bqa
 \widetilde{\mathcal M}^{(\alpha\alpha_s)} &=& \sum_{V_1,V_2=Z,\gamma}
 {-e^2\over s^2-M_{V_1}^2}\bar{v}(k_1)\Gamma_{V_1,\mu}u(k_2)
\bar{u}(p_1)\Gamma_{V_2,\nu}v(p_2)
\nn\\
&\times& {1\over s_{12}-M_{V_2}^2} (-i e) T^{\mu\nu}_{V_1V_2H}(K,P),
\label{NNLO:amplitude}
\eqa
where the sum is extended over $V_1, V_2=Z^0,\gamma$, and
$-i e T^{\mu\nu}_{HV_1V_2}$ is the $HV_1 V_2$ effective vertex,
which depends on $K=k_1+k_2$ and $P=p_1+p_2$.
The gauge boson $V_1$ is coupled with the incoming $e^+e^-$ pair,
whereas the gauge boson $V_2$ is affiliated with the outgoing $\mu^+\mu^-$ pair.
$\Gamma^{\mu}_V$ represents the coupling between the gauge boson and charged leptons,
whose form has already been specified in the paragraph after \eqref{LO:amplitude}.
The electromagnetic coupling of lepton is chiral symmetric, $g_{\gamma}^\pm =1$.

By Lorentz covariance, the renormalized vertex tensor $T^{\mu\nu}_{HV_1V_2}$ can be
decomposed as
\beq
\label{tensor:decomposition}
T^{\mu\nu}_{HV_1V_2}= T_1 {K^{\mu}K^{\nu}\over s}+ T_2 {P^{\mu} P^{\nu}}+
T_3 {K^{\mu} P^{\nu}\over s}+ T_4 {P^{\mu}K^{\nu}\over s}+ T_5 g^{\mu\nu}+
T_6 \epsilon^{\mu\nu\alpha\beta} {K^{\alpha} P^{\beta}\over s},
\eeq
where $T_i (i=1,\cdots,6)$ are Lorentz scalar solely depending on $s$,
$s_{12}$ and $M_{V_{1,2}}^2$.
Furry theorem enforces that $T_6=0$, technically because
$C$-invariance forbids a single $\gamma^5$
to emerge in the trace over the fermionic loop.
Owing to the current conservation associated with
massless leptons, it turns out that only the scalar form factors
$T_{4,5}$ survive in the differential cross sections.

Substituting \eqref{tensor:decomposition} into \eqref{NNLO:amplitude}, utilizing
the factorization scheme \eqref{Factorization:scheme} to
implement the finite $Z^0$ width effect,
we then obtain the rescaled amplitude $\mathcal{M}^{(\alpha\alpha_s)}$.
From \eqref{Higher:order:X:Section:interference}, we find the
${\cal O}(\alpha\alpha_s)$ mixed electroweak-QCD correction to the differential
cross section to be
\bqa
\label{Diff:X:section:NNLO}
{d \sigma^{(\alpha\alpha_s)}\over d s_{12}} &=&\frac{\alpha^3M_Z}{9c_W s_W\sqrt{s}} \left|\mathcal{F}\right|^2 \sum_{V_1,V_2=Z,\gamma}\frac{\left(g^-_{V_1}g^-_Z+g^+_{V_1}g^+_Z\right)
\left(g^-_{V_2}g^-_Z+g^+_{V_2}g^+_Z\right)}{\left(s-M_Z^2\right)\left(s-M_{V_1}^2\right)}
\nn \\
&\times & \frac{s_{12} |{\bf p}_H| }{\left(s_{12}-M_Z^2\right)\left(s_{12}-M_{V_2}^2\right)}
\mathcal{T}_{V_1V_2},
\eqa
where
\beq
\label{Def:cal:T}
\mathcal{T}_{V_1V_2}=  {{\bf p}^2_H \over 2}\left(\frac{1}{s_{12}}-\frac{M_H^2}{s_{12}s}+\frac{1}{s}\right)  T_{4}
+\left({ {\bf p}^2_H\over s_{12} } + 3 \right) T_{5}.
\eeq

\section{Numerical Results}
\label{Numerics}

Following \cite{Sun:2016bel}, we take $\sqrt{s}= 240$, $250$ GeV as two benchmark
CM energies at \textsf{CEPC}. We adopt the following
values for the input parameters~\cite{Patrignani:2016xqp}: $M_H=125.09$ GeV,
$M_Z=91.1876(21)$ GeV, $\Gamma_Z=2.4952(23)$ GeV,
$M_W=80.385(15)$ GeV, $m_t=174.2\pm 1.4$ GeV,
$G_\mu=1.166 3787(6)\times 10^{-5}\;\text{GeV}^{-2}$, $\alpha(0)=1/137.035999$,
$\Delta\alpha^{(5)}_{\text{had}}= 0.02764(13)$, and $\alpha(M_Z)=1/128.943$ in the $\alpha(M_Z)$ scheme.
To analyze the mixed electroweak-QCD correction, we take $\alpha_s(M_Z)=0.1185$,
and use the package \textsf{RunDec}~\cite{Chetyrkin:2000yt}
to evaluate the QCD running coupling constant at other scales.

\begin{figure}[htbp]
\centering
\includegraphics*[width=.5\textwidth]{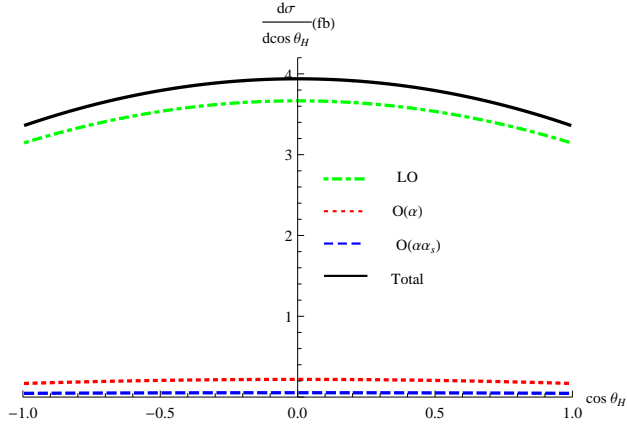}
\caption{Angular distribution of the Higgs boson at $\sqrt{s}=240$ GeV,
shown at various levels of perturbative accuracy.
\label{Fig:angular:distr}}
\end{figure}

\begin{figure}[htbp]
\centering
\includegraphics*[width=.5\textwidth]{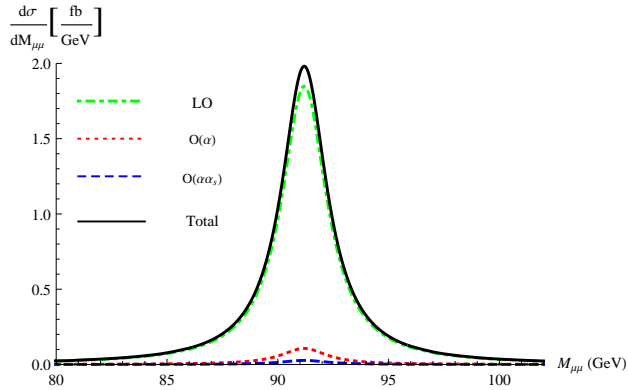}
\caption{$\mu^+\mu^-$ invariant mass spectrum at $\sqrt{s}=240$ GeV, at various levels of perturbative accuracy.
\label{Fig:mumu:mass:distr}}
\end{figure}

The angular distribution of Higgs boson at $\sqrt{s}= 240$ GeV is depicted in Fig.~\ref{Fig:angular:distr},
including both NLO weak and mixed electroweak-QCD corrections.
We stay with the $\alpha(0)$ scheme,  and fix $\mu=\sqrt{s}/2$ for the QCD coupling,
and take $\alpha_s(\sqrt{s}/2)=0.1135$.
The impact of the ${\cal O}(\alpha)$ and ${\cal O}(\alpha\alpha_s)$ corrections to the Higgs
angular distribution is quite analogous to what is found in our preceding work on $e^+e^- \to ZH$~\cite{Sun:2016bel}.
In Fig.~\ref{Fig:mumu:mass:distr}, we also show the $\mu^+\mu^-$ invariant mass spectrum
at $\sqrt{s}=240$ GeV, including both NLO weak and mixed electroweak-QCD corrections.
As expected, the spectrum develops a sharp Breit-Wigner peak around the $Z$ resonance.
The ${\cal O}(\alpha)$ and ${\cal O}(\alpha\alpha_s)$ corrections play a very minor role except
in the proximity of the $Z$ pole.
It is interesting for the future measurement of the di-muon spectrum at \textsf{CEPC}
to examine our predictions.

\begin{table*}
\begin{center}

\begin{tabular}{|c|c|c|c|c|c|c|c|c|c|c|c|c|}

\hline

\multicolumn{2}{|c|}{\backslashbox{$s {d\sigma}/{d s_{12}}$ (fb)}{$ M_{\mu\mu}({\rm GeV})$ }} & 50 & 70 & 80 & 85 & 90 & 91&92 & 95& 100 & 110 & $\sigma$ \\

\hline

\multicolumn{2}{|c|}{LO (fb)} & 0.66 & 2.39& 8.03& 24.45& 309.02& 570.98&407.45& 53.27& 9.66& 1.31& 6.9828\\

\hline

${\mathcal O}(\alpha)$ &Resonant\;(fb)&0.04 &0.14&0.47&1.42&17.78&32.82&23.39&3.05&0.55&0.07& 0.4015\\

\cline{2-13}

  &Nonresonant ($10^{-4}$ fb)&65 &39&22&12&1&0&-0&-7&-16&-24& 8.5\\

  \hline

\multicolumn{2}{|c|}{${\mathcal O}(\alpha\alpha_s)$\;(fb)}&0.01&0.04&0.13&0.35&4.54&8.37&5.97&0.79&0.15&0.02& 0.103\\

\hline

\end{tabular}

\end{center}

\caption{\label{TABLE:1} Differential cross section with respect to the $\mu^+\mu^-$ invariant mass at $\sqrt{s}=240$ GeV.
Note the upper bound for $M_{\mu\mu}$ equals $\sqrt{s}-M_H$.}
\end{table*}

In Table~\ref{TABLE:1}, we supplement more details for the dimuon invariant mass spectrum.
We divide the NLO weak correction into the contribution from the resonant diagrams and
the one from non-resonant diagrams. As can be seen from the Table~\ref{TABLE:1},
the ${\cal O}(\alpha)$ correction is saturated by the resonant diagrams almost in the entire
energy range, especially near the $Z$ peak.

\begin{table}[htbp]
	\renewcommand\arraystretch{1.2}
	\begin{tabular}{|c|c|c|c|c|}
		\hline
		$ \sqrt{s}$ (GeV) & schemes &  $\sigma_{\rm LO}$ (fb) & $\sigma_{\rm NLO}$ (fb) & $\sigma_{\rm NNLO}$ (fb)
		\\
		\hline
		&$ \alpha(0) $& $6.983_{-0.023}^{+0.023}$ & $7.385_{-0.037}^{+0.037}$& $7.488_{-0.036-0.009}^{+0.036+0.004}$
		\\
		\cline{2-5}
		240&$ \alpha(M_Z) $& $8.382_{-0.027}^{+0.028}$ & $7.317_{-0.036}^{+0.037}$ & $7.448_{-0.035-0.011}^{+0.036+0.005}$
		\\
		\cline{2-5}
		&$ G_{\mu} $& $7.772_{-0.004}^{+0.004}$ & $7.527_{-0.017}^{+0.016}$ & $7.554_{-0.017-0.002}^{+0.017+0.001}$
		\\
		\hline
		\hline
		&$ \alpha(0) $& $7.036_{-0.023}^{+0.023}$ & $7.424_{-0.037}^{+0.037}$ & $7.527_{-0.037-0.009}^{+0.037+0.005}$
		\\
		\cline{2-5}
		250&$ \alpha(M_Z) $& $8.446_{-0.028}^{+0.028}$ & $7.350_{-0.036}^{+0.037}$ & $7.481_{-0.037-0.011}^{+0.037+0.006}$
	\\
\cline{2-5}
&$ G_{\mu} $& $7.831_{-0.004}^{+0.004}$ & $7.564_{-0.017}^{+0.017}$ & $7.591_{-0.016-0.002}^{+0.017+0.001}$
\\
\hline
\end{tabular}
\caption{\label{TABLE:2} The total cross section for $e^+e^-\to \mu^+\mu^-H$
at $\sqrt{s}=240\,(250)$ GeV. The LO, NLO, and NNLO predecitions are presented with
three renormalization sub-schemes. To estimate the parametric uncertainty, we take
$M_W=80.385\pm 0.015$ GeV, $m_t=174.2\pm 1.4$ GeV, and
$\Delta\alpha^{(5)}_{\text{had}}= 0.02764\pm 0.00013$.
We also vary the QCD coupling constant from $\alpha_s(M_Z)$ to $\alpha_s(\sqrt{s})$,
with the central value taken as $\alpha_s(\sqrt{s}/2)$.}
\end{table}

Our goal is to present to date the most comprehensive predictions for
the $e^+e^-\to \mu^+\mu^- H$ process, taking into various sorts of theoretical uncertainties account.
In Table~\ref{TABLE:2}, we present our LO, NLO, NNLO predictions for the integrated
cross section at $\sqrt{s}=240\,(250)$ GeV. The results are provided with three
renormalization sub-schemes. We also include the uncertainty inherent in the input parameters
(first error) and the uncertainty due to the QCD renormalization scale (second error).
To assess the parametric uncertainty, we vary the values of $M_W$ and $m_t$, and $\Delta\alpha^{(5)}_{\text{had}}$
around the central PDG values within the $1\sigma$ bands.
For the QCD scale uncertainty, we slide the $\mu$ in $\alpha_s$ from $M_Z$ to $\sqrt{s}$.

From Table~\ref{TABLE:2}, we observe a very similar pattern of scheme and parametric
dependence of higher-order corrections as \cite{Sun:2016bel}.
While the parametric and scale uncertainties of the
NNLO predictions in the $\alpha(0)$ and $\alpha(M_Z)$
schemes are both about 0.5\% of the NNLO results,
the relative errors are somewhat reduced in the $G_\mu$ scheme ($\approx 0.2\%$).
We also find that in the $G_\mu$ scheme,
the mixed electroweak-QCD corrections only amount to
0.4\% of LO cross section, which might be attributed to the fact that in addition to the
running of $\alpha$, universal corrections to the $\rho$
parameter are also absorbed into the LO cross section.
As can also be seen in Table~\ref{TABLE:2}, though the predicted LO cross sections
from three renormalization schemes differ significantly,
including the NLO weak correction significantly help them converge to
each other. Including mixed electroweak-QCD correction appears not to further
reduce the scheme dependence. To yield a scheme-insensitive prediction,
it appears to be imperative to continue to compute the NNLO electroweak correction,
which is certainly an extremely daunting task.

Since $\Gamma_Z\ll M_Z$, and the production rate is predominantly
saturated by the $Z^0$ resonance.
It may seem natural to anticipate that the NWA remains valid even after including higher order corrections.
Under the assumption of NWA, one may approximate the LO cross section and the higher-order radiative
corrections by
\begin{subequations}
\bqa
&& \sigma_0 \big |_{\rm NWA}= \sigma_0(ZH) {\rm Br}_0 (Z\to \mu^+\mu^-),
\\
&& \sigma^{(\alpha)}  \big |_{\rm NWA}= \sigma^{(\alpha)}(ZH) {\rm Br}_0 (Z\to \mu^+\mu^-)
+\sigma_0(ZH) {\rm Br}^{(\alpha)} (Z\to \mu^+\mu^-),
\\
&& \sigma^{(\alpha\alpha_s)} \big |_{\rm NWA}= \sigma^{(\alpha\alpha_s)}(ZH) {\rm Br}_0 (Z\to \mu^+\mu^-)
+\sigma_0(ZH) {\rm Br}^{(\alpha\alpha_s)} (Z\to \mu^+\mu^-),
\eqa
\end{subequations}
where $\sigma(ZH)$ represents the Higgsstrahlung cross section,
with $\sigma_0(ZH)$ given in \eqref{Unpol:LO:ZH:X:section}.
${\rm Br}_0$ is defined in \eqref{Born:Order:Z:mumu:partial:width:Bran:Frac}, and
the radiative corrections ${\rm Br}^{(\alpha\alpha_s^n)}$ ($n=0,1$) can be read off from
\begin{align}
{\rm Br}(Z\to \mu^+\mu^-)&={\rm Br}_0 (Z\to \mu^+\mu^-)+{\rm Br}^{(\alpha)}
(Z\to \mu^+\mu^-)+{\rm Br}^{(\alpha\alpha_s)} (Z\to \mu^+\mu^-)+\cdots.
\end{align}
Since the width of the $Z^0$ is held fixed, the perturbative expansion for the
branching fraction of $Z^0\to \mu^+\mu^-$
amounts to the expansion for the corresponding partial width.

\begin{table}[h!]
\begin{tabular}{|c|c|c|c|}
\hline
&LO & NLO& NNLO \\
\hline
$\sigma$ (fb)&6.983&7.385&7.488\\
\hline
$\sigma|_{\rm NWA}$ (fb)&7.241&7.657&7.760\\
\hline
\end{tabular}
\caption{\label{Table:3} Compare the full and NWA predictions to the cross sections at $\sqrt{s}= 240$ GeV, at various
levels of perturbative accuracy.}
\end{table}

In Table~\ref{Table:3}, we compare the predicted $e^+e^-\to \mu^+\mu^-H$ cross section
from the literal full calculation with that from NWA.
For the sake of concreteness, we take $\sqrt{s}= 240$ GeV, and employ the $\alpha(0)$ scheme.
At LO, the NWA prediction is about $3\%$ higher than the full prediction, while ${\cal O}(\alpha)$ and
${\cal O}(\alpha\alpha_s)$ corrections are observed to be only slightly different.
As a consequence, the NWA prediction to the total cross section at NNLO accuracy turns out to be
about 4\% higher than the full NNLO prediction.

\section{Summary}
\label{Conclusion}

Higgsstrahlung is the leading Higgs production mechanism at \textsf{CEPC}.
The mixed electroweak-QCD correction to $e^+e^-\to ZH$ has recently become available~\cite{Gong:2016jys,Sun:2016bel}.
This piece of NNLO correction appears to be surprisingly large,
about 1\% of the Born-order result, therefore must be considered when matching the
exquisite experimental accuracy.

To make closer contact with the actual experimental measurement, in this work we have investigated both NLO weak
and mixed electroweak-QCD corrections to one of the golden mode in \textsf{CEPC},
{\it i.e.} $e^+e^-\to \mu^+\mu^-H$, with the finite $Z^0$  width properly accounted.
At $\sqrt{s}\approx 240$ GeV, the NLO weak correction may reach $6\%$ of the Born order cross section,
while the NNLO mixed electroweak-QCD correction can reach $1.5\%$ of the LO cross section,
greater than the projected experimental accuracy of $0.9\%$.
We also present numerical predictions to various differential cross sections
at NNLO accuracy, in particular we predict the $\mu^+\mu^-$
invariant-mass spectrum of the Breit-Wigner shape.
We have also compared our full predictions with those based on the NWA,
and found the agreement within a few percents.
It is interesting to await the future experiment to examine our predictions.

We also carefully address the issue about scheme-dependence of our predictions,
at various levels of perturbative accuracy.
Employing three popular renormalization sub-schemes, we find that the predicted
LO cross sections substantially differ from each other.
Including the NLO weak correction is crucial to stabilize the predictions from different schemes,
however including mixed electroweak-QCD correction seems not to help.
To yield a scheme-insensitive prediction,
it appears to be compulsory to continue to include the NNLO electroweak correction,

\vspace{0.2 cm}
\begin{acknowledgments}
We are grateful to Gang Li and Qing-Feng Sun for useful discussions.
The work of W. C. and Y. J. is supported in part by the National Natural Science Foundation of China under Grants
No.~11475188, No.~1187051300, No.~11261130311 (CRC110 by DGF and NSFC).
The work of F. F. is supported by the National Natural Science Foundation of China under Grant No.~11505285, No.~11875318,
and by the Fundamental Research Funds for the Central Universities.
W.-L. S. is supported by the National Natural Science Foundation of China under Grants No.~11605144.
\end{acknowledgments}



\begin{thebibliography}{99}
\bibitem{Aad:2012tfa}
  G.~Aad {\it et al.} [ATLAS Collaboration],
  Phys.\ Lett.\ B {\bf 716}, 1 (2012)
  [arXiv:1207.7214 [hep-ex]].



\bibitem{Chatrchyan:2012xdj}
  S.~Chatrchyan {\it et al.} [CMS Collaboration],
  Phys.\ Lett.\ B {\bf 716}, 30 (2012)
  [arXiv:1207.7235 [hep-ex]].



\bibitem{Accomando:1997wt}
  E.~Accomando {\it et al.} [ECFA/DESY LC Physics Working Group],
  Phys.\ Rept.\  {\bf 299}, 1 (1998)
  [hep-ph/9705442].



\bibitem{Baer:2013cma}
  H.~Baer {\it et al.},
  arXiv:1306.6352 [hep-ph].



\bibitem{Gomez-Ceballos:2013zzn}
  M.~Bicer {\it et al.} [TLEP Design Study Working Group],
  JHEP {\bf 1401}, 164 (2014)
  [arXiv:1308.6176 [hep-ex]].



\bibitem{CEPC-SPPCStudyGroup:2015csa}
  CEPC-SPPC Study Group,
  IHEP-CEPC-DR-2015-01, IHEP-TH-2015-01, IHEP-EP-2015-01.



\bibitem{Ellis:1975ap}
  J.~R.~Ellis, M.~K.~Gaillard and D.~V.~Nanopoulos,
  Nucl.\ Phys.\ B {\bf 106}, 292 (1976).



\bibitem{Lee:1977eg}
  B.~W.~Lee, C.~Quigg and H.~B.~Thacker,
  Phys.\ Rev.\ D {\bf 16}, 1519 (1977).



\bibitem{Lee:1977yc}
  B.~W.~Lee, C.~Quigg and H.~B.~Thacker,
  Phys.\ Rev.\ Lett.\  {\bf 38}, 883 (1977).



\bibitem{Ioffe:1976sd}
  B.~L.~Ioffe and V.~A.~Khoze,
  Sov.\ J.\ Part.\ Nucl.\  {\bf 9}, 50 (1978)
  [Fiz.\ Elem.\ Chast.\ Atom.\ Yadra {\bf 9}, 118 (1978)].



\bibitem{Fleischer:1982af}
  J.~Fleischer and F.~Jegerlehner,
  Nucl.\ Phys.\ B {\bf 216}, 469 (1983).



\bibitem{Kniehl:1991hk}
  B.~A.~Kniehl,
  Z.\ Phys.\ C {\bf 55}, 605 (1992).


\bibitem{Denner:1992bc}
  A.~Denner, J.~Kublbeck, R.~Mertig and M.~Bohm,
  Z.\ Phys.\ C {\bf 56}, 261 (1992).



\bibitem{Gong:2016jys}
  Y.~Gong, Z.~Li, X.~Xu, L.~L.~Yang and X.~Zhao,
  Phys.\ Rev.\ D {\bf 95}, no. 9, 093003 (2017)
  [arXiv:1609.03955 [hep-ph]].



\bibitem{Sun:2016bel}
  Q.~F.~Sun, F.~Feng, Y.~Jia and W.~L.~Sang,
  Phys.\ Rev.\ D {\bf 96}, no. 5, 051301 (2017)
  [arXiv:1609.03995 [hep-ph]].

\bibitem{Mo:2015mza}
  X.~Mo, G.~Li, M.~Q.~Ruan and X.~C.~Lou,
  Chin.\ Phys.\ C {\bf 40}, no. 3, 033001 (2016)
  doi:10.1088/1674-1137/40/3/033001
  [arXiv:1505.01008 [hep-ex]].

\bibitem{Chen:2016zpw}
  Z.~Chen, Y.~Yang, M.~Ruan, D.~Wang, G.~Li, S.~Jin and Y.~Ban,
  Chin.\ Phys.\ C {\bf 41}, no. 2, 023003 (2017)
  doi:10.1088/1674-1137/41/2/023003
  [arXiv:1601.05352 [hep-ex]].

\bibitem{Jones:1979bq}
  D.~R.~T.~Jones and S.~T.~Petcov,
  Phys.\ Lett.\  {\bf 84B}, 440 (1979).



\bibitem{Berends:1984dw}
  F.~A.~Berends and R.~Kleiss,
  Nucl.\ Phys.\ B {\bf 260}, 32 (1985).



\bibitem{Altarelli:1987ue}
  G.~Altarelli, B.~Mele and F.~Pitolli,
  Nucl.\ Phys.\ B {\bf 287}, 205 (1987).



\bibitem{Boos:1993uf}
  E.~Boos, M.~Sachwitz, H.~J.~Schreiber and S.~Shichanin,
  Int.\ J.\ Mod.\ Phys.\ A {\bf 10}, 2067 (1995).



\bibitem{Patrignani:2016xqp}
  C.~Patrignani {\it et al.} [Particle Data Group],
  Chin.\ Phys.\ C {\bf 40}, no. 10, 100001 (2016).



\bibitem{Belanger:2002me}
  G.~Belanger, F.~Boudjema, J.~Fujimoto, T.~Ishikawa, T.~Kaneko, K.~Kato and Y.~Shimizu,
  Nucl.\ Phys.\ Proc.\ Suppl.\  {\bf 116}, 353 (2003)
  [hep-ph/0211268].



\bibitem{Denner:2003yg}
  A.~Denner, S.~Dittmaier, M.~Roth and M.~M.~Weber,
  Phys.\ Lett.\ B {\bf 560}, 196 (2003)
  [hep-ph/0301189].



\bibitem{Denner:2003iy}
  A.~Denner, S.~Dittmaier, M.~Roth and M.~M.~Weber,
  Nucl.\ Phys.\ B {\bf 660}, 289 (2003)
  [hep-ph/0302198].



\bibitem{Belanger:2003sd}
  G.~Belanger, F.~Boudjema, J.~Fujimoto, T.~Ishikawa, T.~Kaneko, K.~Kato and Y.~Shimizu,
  Phys.\ Rept.\  {\bf 430}, 117 (2006)
  [hep-ph/0308080].



\bibitem{Denner:2004jy}
  A.~Denner, S.~Dittmaier, M.~Roth and M.~M.~Weber,
  Nucl.\ Phys.\ Proc.\ Suppl.\  {\bf 135}, 88 (2004)
  [hep-ph/0406335].



\bibitem{Belanger:2002ik}
  G.~Belanger, F.~Boudjema, J.~Fujimoto, T.~Ishikawa, T.~Kaneko, K.~Kato and Y.~Shimizu,
  Phys.\ Lett.\ B {\bf 559}, 252 (2003)
  [hep-ph/0212261].



\bibitem{Boudjema:2004eb}
  F.~Boudjema, J.~Fujimoto, T.~Ishikawa, T.~Kaneko, K.~Kato, Y.~Kurihara, Y.~Shimizu and Y.~Yasui,
  Phys.\ Lett.\ B {\bf 600}, 65 (2004)
  [hep-ph/0407065].



\bibitem{Kilian:2007gr}
  W.~Kilian, T.~Ohl and J.~Reuter,
  Eur.\ Phys.\ J.\ C {\bf 71}, 1742 (2011)
  [arXiv:0708.4233 [hep-ph]].



\bibitem{Veltman:1963th}
  M.~J.~G.~Veltman,
  Physica {\bf 29}, 186 (1963).



\bibitem{Stuart:1991xk}
  R.~G.~Stuart,
  Phys.\ Lett.\ B {\bf 262}, 113 (1991).



\bibitem{Sirlin:1991fd}
  A.~Sirlin,
  Phys.\ Rev.\ Lett.\  {\bf 67}, 2127 (1991).



\bibitem{Baur:1991pp}
  U.~Baur, J.~A.~M.~Vermaseren and D.~Zeppenfeld,
  Nucl.\ Phys.\ B {\bf 375}, 3 (1992).



\bibitem{Kurihara:1994fz}
  Y.~Kurihara, D.~Perret-Gallix and Y.~Shimizu,
  Phys.\ Lett.\ B {\bf 349}, 367 (1995)
  [hep-ph/9412215].



\bibitem{Beenakker:1996kt}
  W.~Beenakker {\it et al.},
  hep-ph/9602351.



\bibitem{Beenakker:1996kn}
  W.~Beenakker, G.~J.~van Oldenborgh, A.~Denner, S.~Dittmaier, J.~Hoogland, R.~Kleiss, C.~G.~Papadopoulos and G.~Passarino,
  Nucl.\ Phys.\ B {\bf 500}, 255 (1997)
  [hep-ph/9612260].

\bibitem{Beuthe:1996fe}
  M.~Beuthe, R.~Gonzalez Felipe, G.~Lopez Castro and J.~Pestieau,
  Nucl.\ Phys.\ B {\bf 498}, 55 (1997)
  doi:10.1016/S0550-3213(97)00263-0
  [hep-ph/9611434].

\bibitem{Denner:1999gp}
  A.~Denner, S.~Dittmaier, M.~Roth and D.~Wackeroth,
  Nucl.\ Phys.\ B {\bf 560}, 33 (1999)
  [hep-ph/9904472].



\bibitem{Denner:2006ic}
  A.~Denner and S.~Dittmaier,
  Nucl.\ Phys.\ Proc.\ Suppl.\  {\bf 160}, 22 (2006)
  [hep-ph/0605312].



\bibitem{Beneke:2003xh}
  M.~Beneke, A.~P.~Chapovsky, A.~Signer and G.~Zanderighi,
  Phys.\ Rev.\ Lett.\  {\bf 93}, 011602 (2004)
  [hep-ph/0312331].



\bibitem{Beneke:2015vfa}
  M.~Beneke,
  Nucl.\ Part.\ Phys.\ Proc.\  {\bf 261-262}, 218 (2015)
  [arXiv:1501.07370 [hep-ph]].



\bibitem{Willenbrock:1991hu}
  S.~Willenbrock and G.~Valencia,
  Phys.\ Lett.\ B {\bf 259}, 373 (1991).



\bibitem{Hahn:2000kx}
  T.~Hahn,
  Comput.\ Phys.\ Commun.\  {\bf 140}, 418 (2001)
  [hep-ph/0012260].



\bibitem{Mertig:1990an}
  R.~Mertig, M.~Bohm and A.~Denner,
  Comput.\ Phys.\ Commun.\  {\bf 64}, 345 (1991).



\bibitem{Feng:2012tk}
  F.~Feng and R.~Mertig,
  arXiv:1212.3522 [hep-ph].



\bibitem{Shtabovenko:2016sxi}
  V.~Shtabovenko, R.~Mertig and F.~Orellana,
  Comput.\ Phys.\ Commun.\  {\bf 207}, 432 (2016)
  [arXiv:1601.01167 [hep-ph]].



\bibitem{Denner:2016kdg}
  A.~Denner, S.~Dittmaier and L.~Hofer,
  Comput.\ Phys.\ Commun.\  {\bf 212}, 220 (2017)
  [arXiv:1604.06792 [hep-ph]].



\bibitem{Hahn:1998yk}
  T.~Hahn and M.~Perez-Victoria,
  Comput.\ Phys.\ Commun.\  {\bf 118}, 153 (1999)
  [hep-ph/9807565].



\bibitem{Denner:1991kt}
  A.~Denner,
  Fortsch.\ Phys.\  {\bf 41}, 307 (1993)
  [arXiv:0709.1075 [hep-ph]].



\bibitem{Feng:2012iq}
  F.~Feng,
  Comput.\ Phys.\ Commun.\  {\bf 183}, 2158 (2012)
  [arXiv:1204.2314 [hep-ph]].



\bibitem{Smirnov:2014hma}
  A.~V.~Smirnov,
  Comput.\ Phys.\ Commun.\  {\bf 189}, 182 (2015)
  [arXiv:1408.2372 [hep-ph]].



\bibitem{Smirnov:2015mct}
  A.~V.~Smirnov,
  Comput.\ Phys.\ Commun.\  {\bf 204}, 189 (2016)
  [arXiv:1511.03614 [hep-ph]].



\bibitem{CubPack}
R.~Cools and A.~Haegemans,
ACM Trans. Math. Softw. {\bf 29} (2003), no.~3~287~C296.


\bibitem{Chetyrkin:2000yt}
  K.~G.~Chetyrkin, J.~H.~Kuhn and M.~Steinhauser,
  Comput.\ Phys.\ Commun.\  {\bf 133}, 43 (2000)
  [hep-ph/0004189].

\end{thebibliography}
\end{document}